\documentclass[twocolumn]{aa}
\usepackage{graphicx}
\usepackage{txfonts}
\begin{document}
   \title{Organic matter in Seyfert2 nuclei :
comparison with our Galactic center lines of sight.   
\thanks{Based on observations collected at the European Southern Observatory, Chile and observations with ISO, an ESA project with instruments funded by ESA Member States.} }

   \author{
$\;$
E. Dartois\inst{1}
          \and
          O. Marco\inst{2}
          \and
          G. M. Mu\~noz-Caro \inst{1}
          \and
          K. Brooks \inst{3,2}
          \and
          D. Deboffle \inst{1}
          \and
          L. d'Hendecourt \inst{1}
          }

   \offprints{E. Dartois}

   \institute{Institut d'Astrophysique Spatiale - UMR-8617
 Université Paris-Sud, bâtiment 121
 F-91405 Orsay, France\\
              \email{emmanuel.dartois@ias.u-psud.fr}
         \and
             European Southern Observatory, Alonso de Cordova 3107, Vitacura, Santiago, Chile\\
             \email{o.marco@eso.org}
         \and
             Departamento de Astronomía, Universidad de Chile, Casilla 36-D, Santiago, Chile
             }

%   \date{Received , 2004; accepted 2004}

   \abstract{
We present ESO - Very Large Telescope and ESA - Infrared Space
Observatory 3 to 4~$\mu$m spectra of Seyfert 2 nuclei as compared to
our galactic center lines of sight. The diffuse interstellar medium
probed in both environments displays the characteristic 3.4~$\mu$m
aliphatic CH stretch absorptions of refractory carbonaceous
material. The profile of this absorption feature is similar in all
sources, indicating the CH$_2$/CH$_3$ ratios of the carbon chains
present in the refractory components of the grains are the same in
Seyfert 2 inner regions. At longer wavelengths the circumstellar
contamination of most of the galactic lines of sight precludes the
identification of other absorption bands arising from the groups
constitutive of the aliphatics seen at 3.4~$\mu$m. The clearer
continuum produced by the Seyfert~2 nuclei represents promising lines
of sight to constrain the existence or absence of strongly infrared
active chemical groups such as the carbonyl one, important to understand
the role of oxygen insertion in interstellar grains. The Spitzer Space
Telescope spectrometer will soon allow one to investigate the
importance of aliphatics on a much larger extragalactic sample.
   \keywords{ISM : dust, extinction, lines and bands - Galaxies : Seyfert 2 - Individual objects :Sgr~A*, GC IRS7, NGC~1068, NGC~7172, IRAS~19254-7245, NGC~5506} }

   \maketitle
%
%________________________________________________________________

\section{Introduction}
Since the first observations (e.g. \cite{1981Natur.294..239A}) two
decades ago, the comparison between infrared features arising from
solid state matter, or dust, observed toward many galactic lines of
sight, and the studies in the laboratory of materials produced under
simulated interstellar conditions has provided some insight into this
important component of our Galaxy, which reflects not only its
chemical evolution but also radiative transfer effects in many
complicated regions. Infrared dust absorption features observed in the
Diffuse InterStellar Medium (DISM) and molecular clouds (MC) allow us to
infer the solid state matter composition, its evolution, and thus
follow the cycling of matter in the Galaxy. Silicates, responsible for
the features at 10~$\mu$m (Si-O stretching mode) and 18~$\mu$m (Si-O
bending mode), as well as carbonaceous material, giving rise to
absorption bands around 3.4~$\mu$m (\cite{2002ApJS..138...75P};
\cite{2000ApJ...537..749C};
\cite{1994ApJ...437..683P}; \cite{1991ApJ...371..607S}) and PAH
emission lines (3.3, 6.2, 7.7, 8.6, 11.3~$\mu$m,
\cite{1984A&A...137L...5L}), dominate the composition of refractory
dust particles present in the DISM.  The relation between emission and
absorption carbon based observed features has important implications
for the cycling of dust in galaxies (\cite{1989irsa.rept..189L}).\\

The exact composition of the carbon-dominated component is still under
debate.  So far, the identification of the carbonaceous material seen
in absorption has proceeded via the spectroscopic analysis of the
so-called 3.4~$\mu$m absorption features due to -CH$_2$- and -CH$_3$-
aliphatic stretching vibrations in hydrocarbon chains.  This 3.4~$\mu$m
absorption is only clearly observed in lines of sight toward our
Galactic center (e.g. \cite{2002ApJ...570..198C} and references
therein), or more recently in the foreground diffuse medium toward a
luminous Young Stellar Object (YSO) on the shoulder of a strong
circumstellar water ice mantle absorption feature
(\cite{2002AJ....124.2790I}). In all these lines of sight, it is clear
that, in addition to the diffuse interstellar component, high amounts
of molecular cloud material and/or circumstellar features also
contribute to the extinction, rendering the spectral analysis
more complex as one has to decipher the underlying respective
contributions of both intervening MC and DISM media. In particular,
strong ice mantle absorption features from the densest regions probed
by the infrared beam will add several contributions (e.g. 3.1~$\mu$m
water ice, 3.47~$\mu$m hydrate, 4.27~$\mu$m CO$_2$, 4.67~$\mu$m CO,
6~$\mu$m water ice, 6.8~$\mu$m CH$_3$ deformation modes, methanol,
7.7~$\mu$m methane, 13~$\mu$m water ice;
\cite{2002A&A...394.1057D};\cite{2000ApJ...537..749C}). To 
reduce the influence of the MC contribution to the spectra, to perform
a finer spectroscopic analysis of the intrinsic DISM absorptions, it
is essential to be able to probe DISM on larger galactic scales and not
only toward embedded individual infrared sources.\\

Several observations using the Infared Space Observatory (ISO)
spectrometers have allowed the identification of some extragalactic
molecular cloud components (\cite{2002A&A...385.1022S}) or
extragalactic diffuse components (\cite{2000A&A...359..900L}). More
recently, ground-based telescopes in specific atmospheric windows
clearly demonstrated the extragalactic sources we should focus our
attention on. Indeed, the infrared
spectra of some external galaxies such as the Seyfert 2 (Sy2) Active
Galactic Nuclei show a strong 3.4~$\mu$m diffuse medium
absorption feature, with no major contribution from molecular cloud
absorption (e.g. \cite{2000MNRAS.319..331I}).\\

We present in this paper observations of both galactic and Seyfert~2
extragalactic lines of sight from near (3.1-~4$\mu$m) to mid
(5.4-7.4~$\mu$m) infrared to investigate the composition of the
refractory hydrocarbons present in the interstellar grains. We first
present the observations and the observed extragalactic lines of
sights in $\S$2 and $\S$3, then compare in $\S$4 the observed
3.4~$\mu$m absorption profiles, in $\S$5 the carbonaceous to silicates
ratios, and the mid-infrared spectra in $\S$6. We discuss in
$\S$7 the implications for the nature of the DISM hydrocarbons and
conclude.

%__________________________________________________________________

\section{Observations}

The VLT spectra presented here were extracted from the ESO Science
Archive Facility (http://archive.eso.org/) from the programs
(69.A-0643 and 69.B-0101) and observations from the
authors programs (67.B-0332 and 71.B-0404). A summary of the main
parameters are given in Table \ref{Sources}. Data were reduced using
in-house software and classical infrared extraction techniques.

The ISO spectra were extracted from the Infrared Space Observatory
(ISO) database (http://isowww.estec.esa.nl/). The different AOT bands
were stitched together by applying gain factors, using overlapping
regions of the spectra to determine them. The corrections applied were
less than 5\%-10\%, in agreement with the differences of the apertures
and the absolute photometric reliability of the spectrometer.

%__________________________________________________ One column table
   \begin{table*} \caption[]{Source parameters} \label{Sources} $$
   \begin{array}{p{0.15\linewidth}llclllll} \hline \noalign{\smallskip}
   Source Name &Date &Inst/AOT \;^{\rm{a}} &FOV/slit \;^{\rm{b}} &R  &\alpha (2000) &\delta (2000) &z &L Mag\;^{\rm{c}}\\
\hline
Galactic center & & & & & & & &\\
\hline
GC SgrA$*$	&19\;\rm{Feb}\;1996	&\rm{ISO}/\rm{SWS01} &14''\rm{x}20''  &R	&17h45m39.97s 	&-29d00'28.7"  &- &3.1\\
%Observer          :RGENZEL
%Proposal          :MPEXGAL1
%Start Time        :19-Feb-1996 15:27:25
%End Time          :19-Feb-1996 17:16:13
%On Target Time [s]:6528 sec 
GC SgrA$*$        &21\;\rm{Feb}\;1997 &\rm{ISO}/\rm{SWS06}  &14''\rm{x}20'' &R   &17h45m39.97s &-29d00'28.8"  &- &\\
%Observer          :DKUNZE
%Proposal          :GCSPEC
%Start Time        :21-Feb-1997 15:24:27
%End Time          :21-Feb-1997 19:03:01
%On Target Time [s]:13114 sec 
GCS 3 I          &29\;\rm{Aug}\;1996             &\rm{ISO}/\rm{SWS01}  &14''\rm{x}20'' &R &17h46m14.84s &-28d49'33.8"  &- &3.2\\
%Observer          :DWHITTET
%Proposal          :ICE_BAND
%Start Time        :29-Aug-1996 22:13:44
%End Time          :29-Aug-1996 23:11:18
%On Target Time [s]:3454 sec 

GCS 3 I            &08\;\rm{Oct}\;1996        &\rm{ISO}/\rm{SWS06} &14''\rm{x}20'' &R &17h46m14.84s &-28d49'33.7"  &- &\\
%Observer          :DWHITTET
%Proposal          :ICE_BAND
%Start Time        :08-Oct-1996 23:50:50
%End Time          :09-Oct-1996 00:44:36
%On Target Time [s]:3226 sec 
%Quality           :Good
IRAS17424-2859      &01\;\rm{Apr}\;1996          &\rm{ISO}/\rm{SWS01} &14''\rm{x}20'' &R &17h45m39.93s &-29d00'22.8"  &- &5\\
%Observer          :MUIZON
%Proposal          :MMDUSTY1
%Start Time        :01-Apr-1996 07:37:44
%End Time          :01-Apr-1996 08:35:26
%On Target Time [s]:3462 sec 

GC IRS7        &29\;\rm{May}\;2002 &VLT\;ISAAC/LWS3\;LR &0.6'' &600 &17h45m41.1s     &-29d00'44.8''     &- &5\\
\hline
Seyfert sources & & & & & & & &\\
\hline
NGC 1068        &18\;\rm{Aug}\;2001 &VLT\;ISAAC/LWS3\;LR &1.0'' &360 &02h42m40.7s     &-00d00m48s      &0.003793 &5.4\\
NGC 7172        &05\;\rm{Aug}\;2001 &VLT\;ISAAC/LWS3\;LR &1.0'' &360 &22h02m01.7s     &-31d52m18s      &0.008683 &8.5\\
IRAS 19254-7245 &04\;\rm{Jun}\;2002 &VLT\;ISAAC/LWS3\;LR &1.0'' &360 &19h31m21.4s     &-72d39m18s      &0.061709 &11.4\\
NGC 5506        &18\;\rm{Jun}\;2003 &VLT\;ISAAC/LWS3\;LR &1.0'' &360 &14h13m14.87s    &-03d12m27s      &0.006068 &7.4\\
            \hline
         \end{array}
     $$
\begin{list}{}{}
\item (a) AOT : Astronomical Observation Template.
\item (b) FOV : Field Of View, for ISO-SWS data; slit width for VLT-ISAAC ones.
\item (c) given L band magnitude are approximate as they are aperture dependent for most of the sources. ISO-SWS spectra are recorded in a source crowded area with a beam size of about 14''$\times$20''
\end{list}
   \end{table*}

%___________________________________________________

\section{Characteristics of the extragalactic sources}

The commonly-accepted AGN model involves a central engine 
(black hole plus accretion disc)
surrounded by sets of dense matter clouds (the so-called broad-line and narrow-line regions, 
BLR \& NLR) and by a dusty/molecular disc-like and thick structure (also called a torus)
which funnels the emission of high energy photons and particles along privileged directions 
(the ionizing cone, the radio jet axis). This in turn gives rise to 
viewing-angle dependent effects which are
usually invoked to explain the various types of AGN discovered so far
(type 1 with broad H$\alpha$ directly visible or 
type 2 with broad H$\alpha$ not directly visible). 
The spatial extension of the infrared emitting region of Seyfert galaxies
is usually in the range 10--100~pc and corresponds to dust heated by the UV emission
coming from the accretion disk (sized $\leq 0.01$~pc). 
The typical extinction on the line of sight is high, $A_v$=15--50 mag.,
with a substantial column density of X-ray absorbing material ($N_H \sim 10^{22-24} \mbox{cm}^{-2}$).
ISAAC spectroscopy has been done using a slit width of 1 arcsec, centered on the
infrared L-band brightest peak of emission of the sources, corresponding to the hot dust 
(T$\sim$500--1000~K) around the AGN central engine.
This means that the slit used with ISAAC is getting all the flux
from this region, for all sources.

\subsection*{Individual observed sources}

NGC\,1068 is the archetypal Seyfert~2 nucleus, nearby and very bright.
It harbours a hidden nucleus with broad permitted emission lines (BLR)
seen in spectropolarimetry. The optical-UV luminosity of the central
source is $\sim 1.5 \times 10^{45} \mbox{erg s}^{-1}$. Adaptive optics
observations (\cite{2000A&A...353..465M}) have shown that the hot
(T$>$900~K) dust emission seems to be confined in a region of
$\sim$10~pc radius around the core. L-band images show a bright,
unresolved central peak of emission surrounded by extended emission,
with 90\% of the flux contained in a region less than 100~pc away from
the central peak.

NGC\,7172 is a Seyfert~2, it has an obscured ($N_H \sim 8 \times 10^{22} \mbox{cm}^{-2}$)
active nucleus with strong infrared emission from the nucleus 
(m$_K$=10 and $L_{IR} \sim 10^{10.45} L \sun$).

IRAS\,19254-7245, also known as The Superantennae, belongs to the 
Ultra Luminous Infrared Galaxies (ULIRG). Its southern nucleus,
the dominant source at different wavelengths,
is classified as a Seyfert~2 galaxy, with the presence of a Compton-thick
AGN with intrinsic luminosity $10^{44} \mbox{erg s}^{-1}$.
An image observed with ISOCAM at 6.75~$\mu m$ shows an unresolved central
source of an angular size $\leq$3 arcsec, with hot dust.

NGC\,5506 is Seyfert~2 with hot dust obscuring the central source.
A feature of silicate absorption has been detected at 9.7~$\mu m$. 
The mid-infrared continuum of the nucleus of NGC 5506, 
is believed to come from warm ($\sim $300~K) 
dust in thermal equilibrium, which should be located within $\sim $10~pc 
of the AGN in the case of the UV luminosity with $\sim 10^{11}$ L [o].
It has recently been reclassified as an optically-obscured Narrow Line Seyfert~1.

%IRAS\,08572+3915 is a ULIRG classified as a liner, with a
%high obscuration towards the central source, believed to be ascribed to dust.

\begin{figure*}
%   \centering
\begin{minipage}{5.1cm}
   \includegraphics[width=5cm]{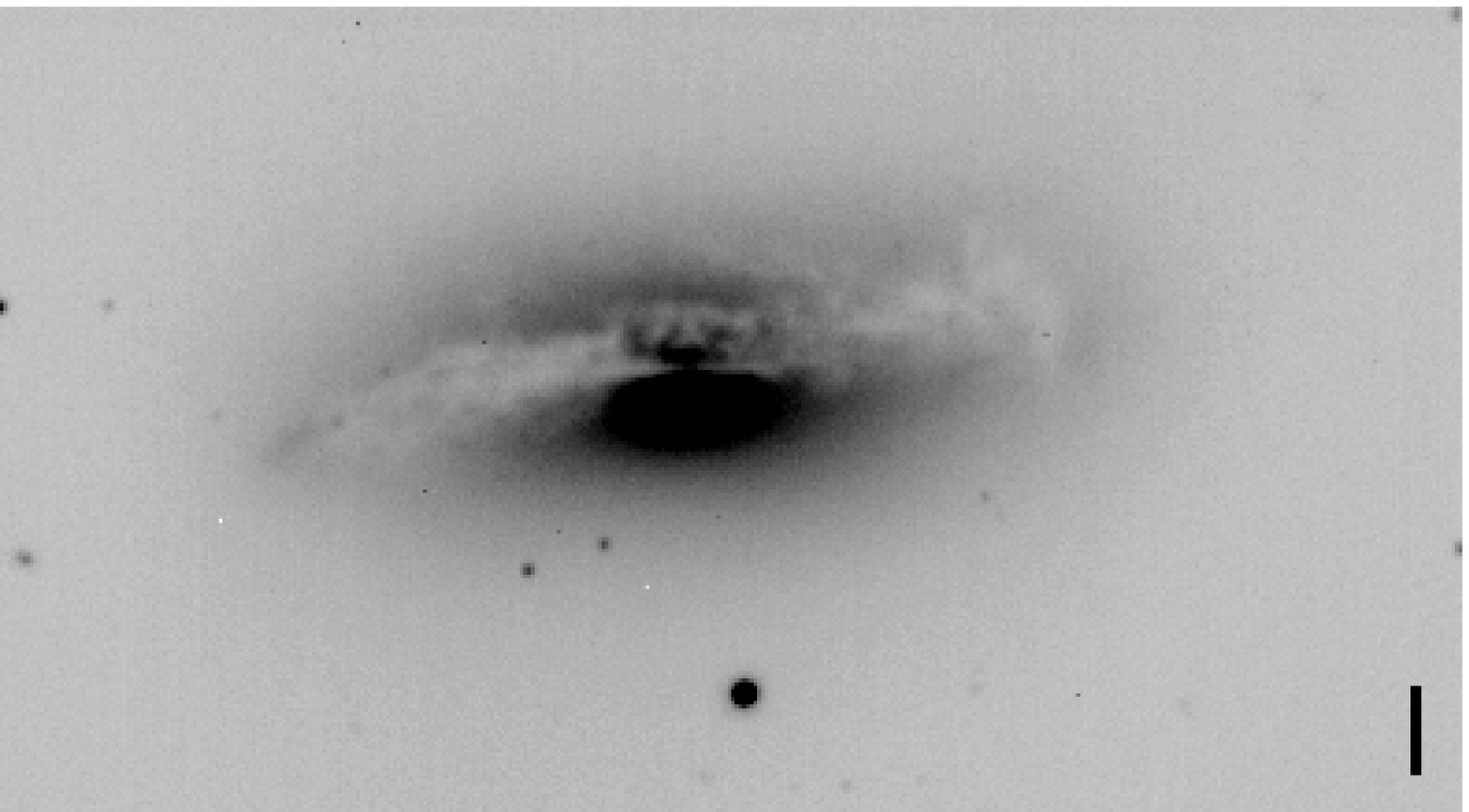}
    \includegraphics[width=7cm]{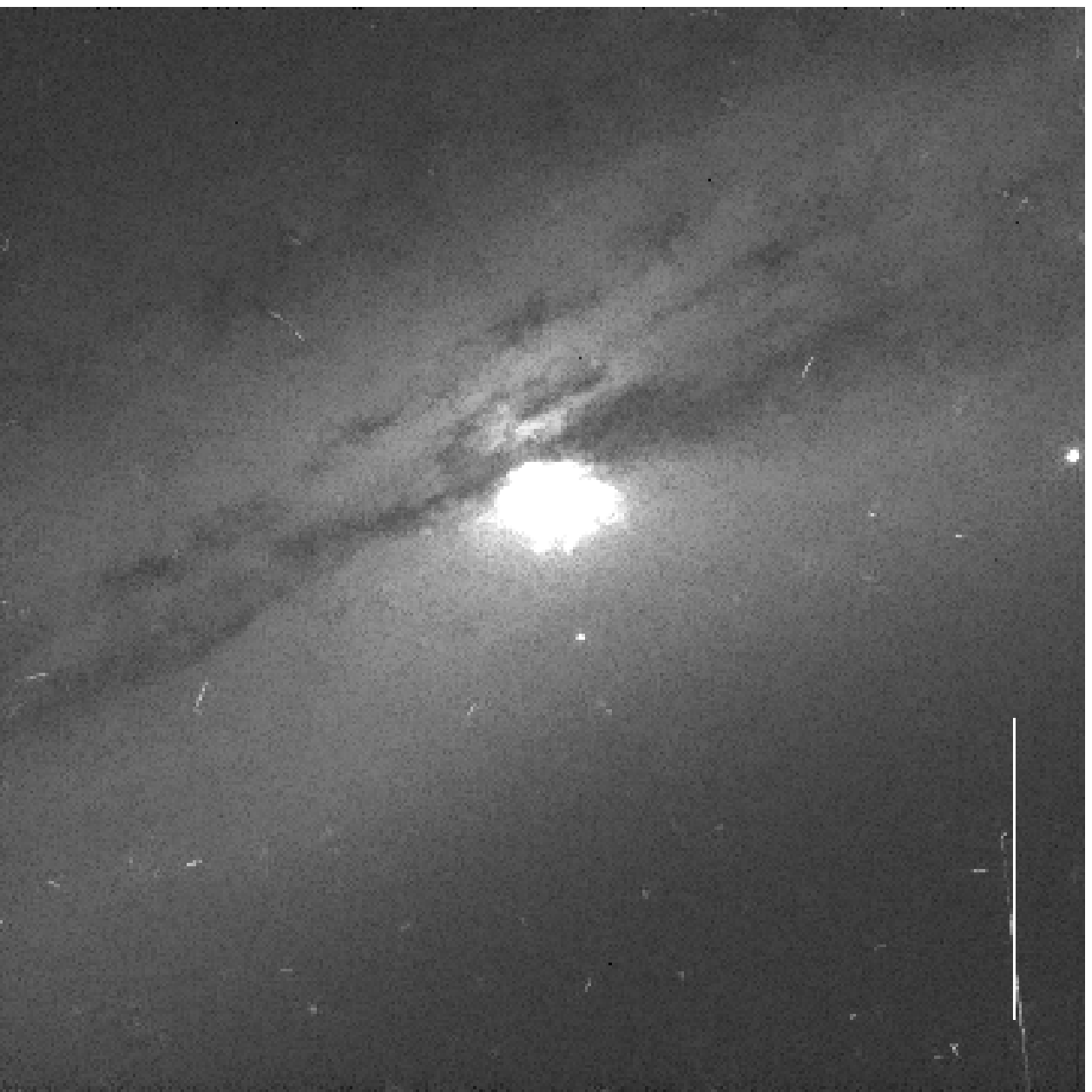}
\end{minipage}
\begin{minipage}{4.5cm}
\vspace*{.8cm}

    \includegraphics[width=5.4cm]{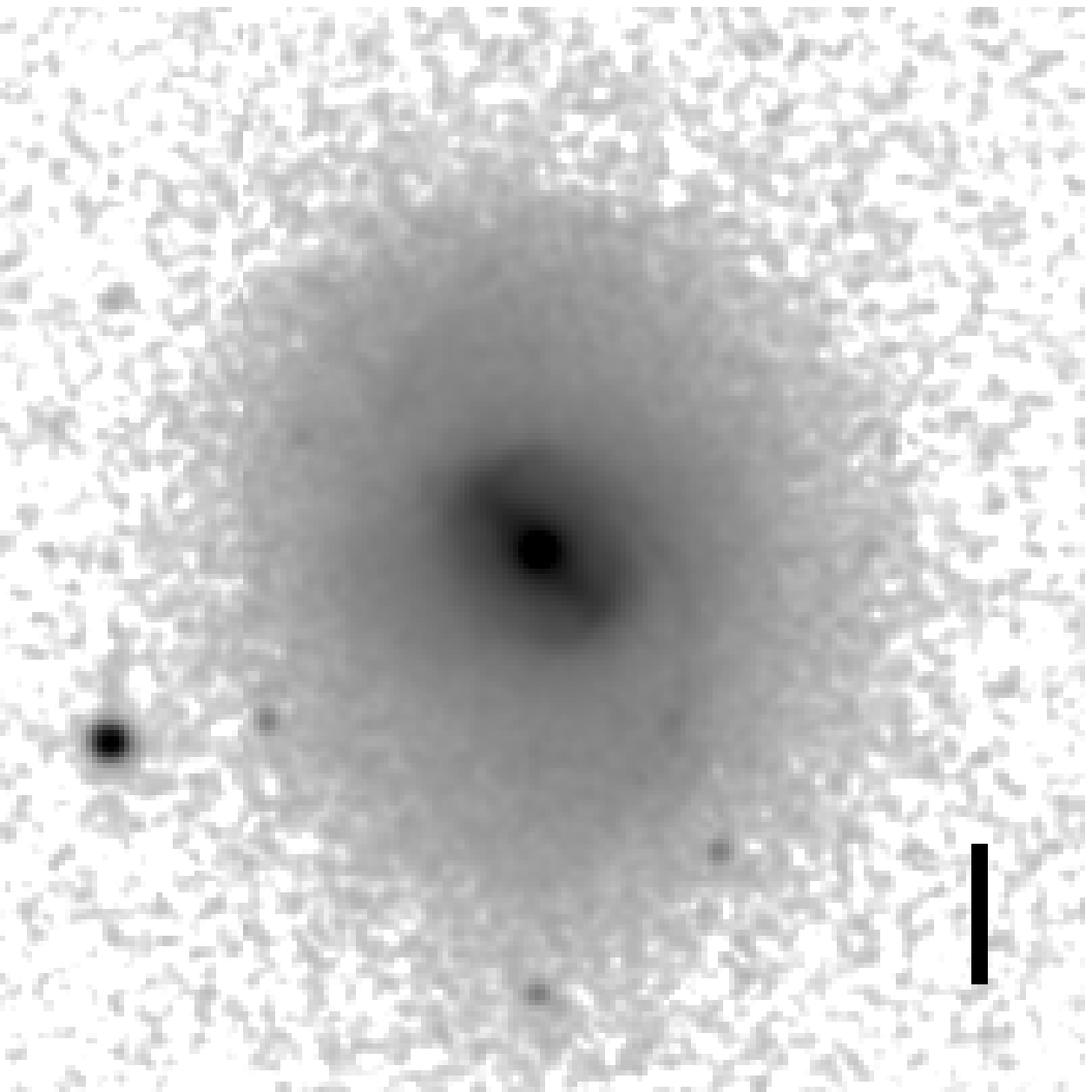}
    \includegraphics[width=5.4cm]{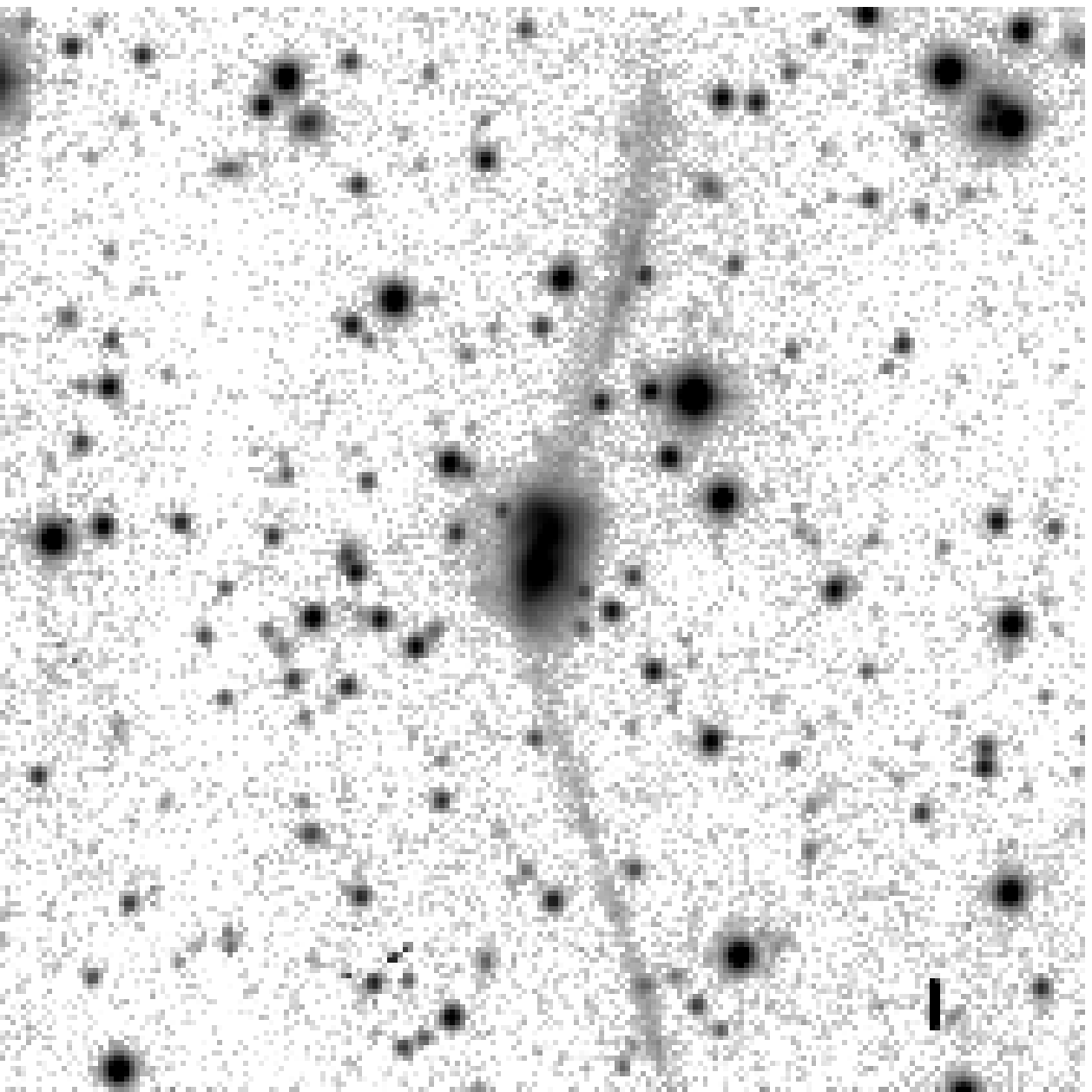}
\end{minipage}
\begin{minipage}{6cm}
   \caption{
Images of the observed sources, showing the inclination of the host
Seyfert 2 galaxies and where dust obscuration is evident, were
retrieved from the NED database
(http://nedwww.ipac.caltech.edu/). From left to right, up to down :
NGC7172 (Canada-France-Hawaii Telescope, \cite{1994hickson}), NGC1068 (2MASS K band,
\cite{2003AJ....125..525J}), NGC5506 (Hubble Space Telescope, 606~nm
filter, \cite{1998ApJS..117...25M}) and IRAS19254-7245 (Danish
Telescope, \cite{2000ApJS..131...71C}). A vertical bar representing a 10'' scale has been drawn.
              \label{fig_images}%
		}
\end{minipage}
    \end{figure*}

Large scale images of the observed extragalactic objects are presented
in Fig.~\ref{fig_images} and their main characteristics in
Table~\ref{agns}.

%______________________________________________ 
%______________________________________________ 

\begin{table*}
\caption{Main characteristics of the extragalactic sources}\label{agns}
\begin{tabular}{lcccc}
\hline
Source name 	&Distance &1'' equiv.    &size of IR emitting region &IR Luminosity \cr
                &Mpc      &pc           &arcsec                     &Lsol (log10)\cr
\hline
NGC 1068	&15       &73           &$<$2.0                     &11.30 $^{\mathrm{a}}$ \cr
NGC 7172	&35       &168          &$<$2.5                     &10.45 $^{\mathrm{b}}$ \cr
IRAS 19254-7245	&240      &1160         &                           &12.06 $^{\mathrm{b}}$ \cr
NGC 5506	&24       &117          &                           &10.36 $^{\mathrm{a}}$ \cr  
%IRAS 08572+3915	&227      &1100         &                           &12.17 \cr
\hline
\end{tabular}
\begin{list}{}{}
\item
(${\mathrm{a}}$) \cite{1999AJ....118.2625R}. (${\mathrm{b}}$) \cite{1997ApJ...486..132B}.
\end{list}
\end{table*}

%______________________________________________ 

\section{3.4~$\mu$m absorption feature}

%______________________________________________ 
   \begin{figure*}
%   \centering
\begin{minipage}{10cm}
\hspace*{-.5cm}
    \includegraphics[width=10cm]{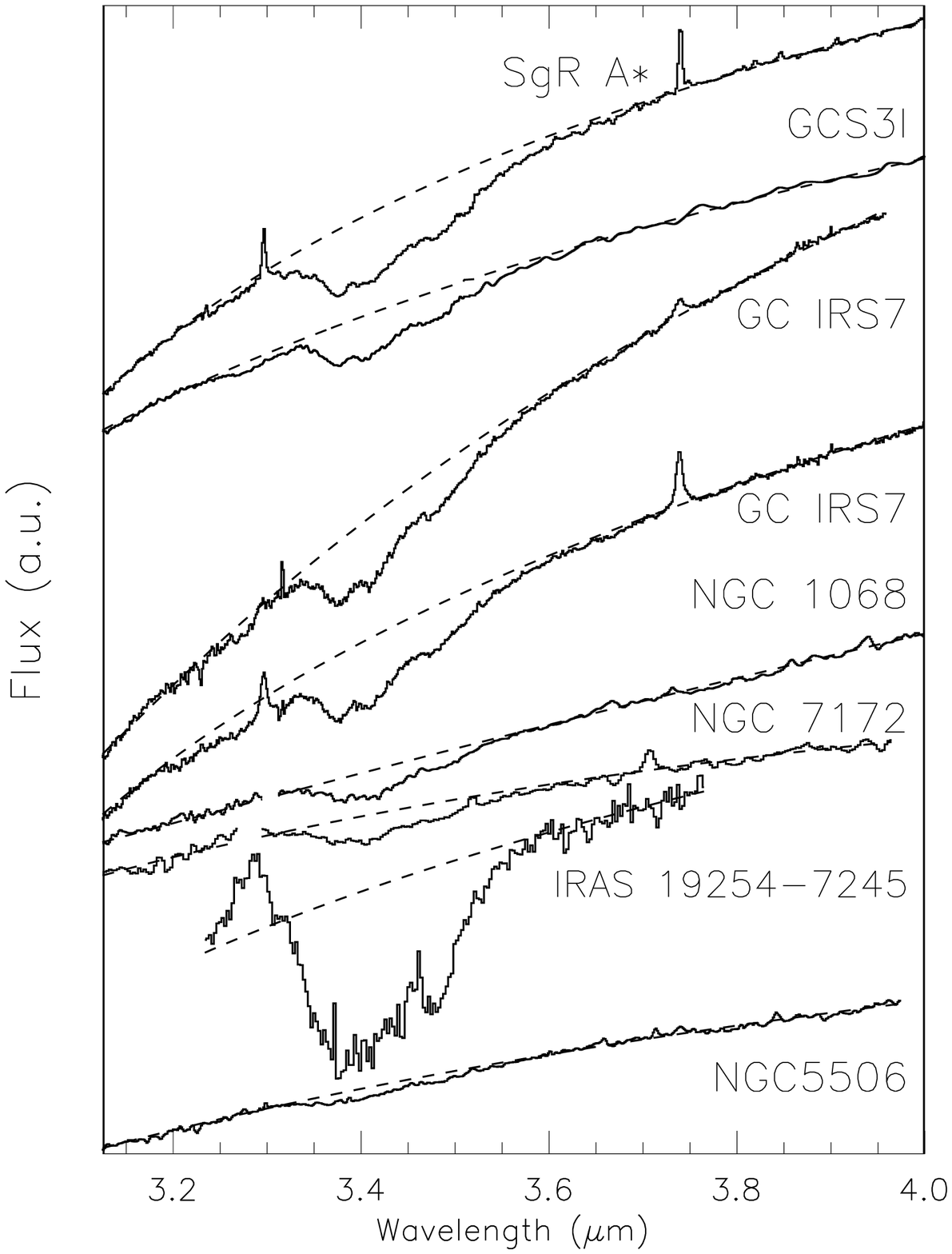}
\end{minipage}
\begin{minipage}{10cm}
\hspace*{-1.5cm}
    \includegraphics[width=10cm]{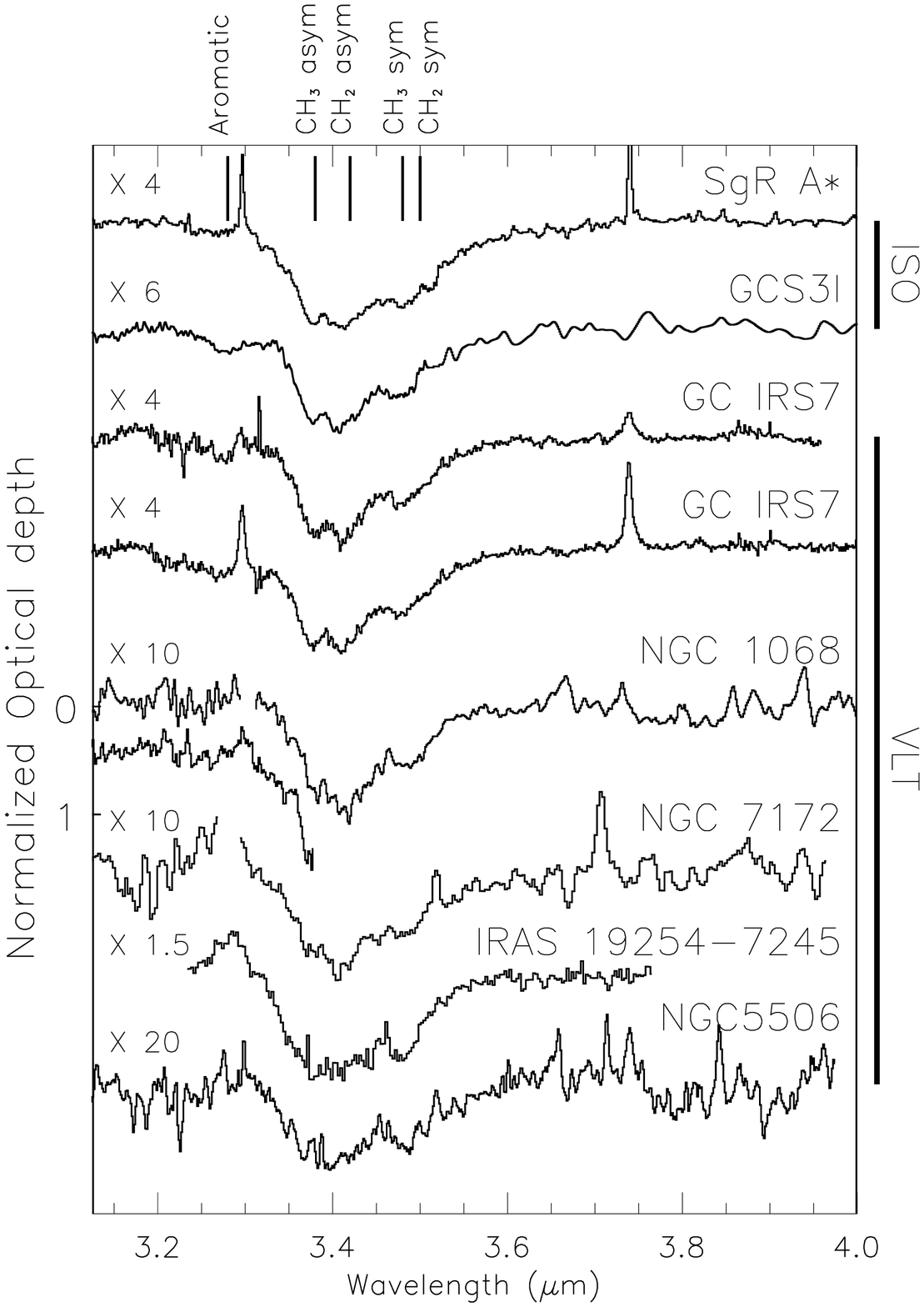}
\end{minipage}
   \caption{
Left : Spectra of the observed sources and adopted local continua to
extract the 3.4~$\mu$m absorption feature, scaled by multiplicative
factors for clarity. The L band magnitudes, corresponding to each
source, are given in Table
\ref{Sources}. The two GC IRS7 independant observations were made with two
different slit orientations on the sky and give rise to slightly
different continua.  $\;\;$ Right : Normalised optical depth spectra
in the 3.15 to 4~$\mu$m of galactic and extragalactic sources. The two
upper traces correspond to ISO Short Wavelength Spectrometer AOT06
spectra (R$\sim$1500), except for GCS3I which was combined above
3.5~$\mu$m with a lower resolution spectrum (SWS01 R$\sim$500). The
other spectra are VLT ISAAC Long Wavelength Low resolution spectra
(R$\sim$360 to 600 in the used modes), except the 3.1-3.35~$\mu$m
SWS06 spectrum of NGC1068 presented below the VLT one. A normalising
factor has been applied to allow the comparison of the absorption
profiles, and is indicated on the left of each spectra. All
extragalactic spectra have been corrected in frequency for their
redshift as given in Table
\ref{Sources}.
              \label{Fig1}%
		}
    \end{figure*}
%______________________________________________ 
%
The 3.4~$\mu$m hydrocarbon absorption features observed in galactic
and extragalactic lines of sight are presented in Fig \ref{Fig1}. A
local continuum has been subtracted from all spectra, including the wing
of a water ice stretching mode absorption band in our Galactic sources
lines of sight, as shown by \cite{2002ApJ...570..198C}. All optical
depth spectra have been multiplied by a factor, indicated on the left,
in order to be able to compare the intrinsic profiles of the bands
with an approximately equal optical depth.

The profile observed toward SgR~A$*$ with ISO appears peculiar as
compared to the others, with more sub-structures.  This can be
explained by the mixing of many infrared emitting sources filling the
large observation beam (about 10$''$) of ISO, leading to some
confusion in this crowded line of sight (see
\cite{2003Msngr.113...17E} for a recent infrared image of this area),
even if there is a dominating infrared source.
The two independent GC~IRS7 VLT observations, with different
orientations of the slit in the sky, better sample the Diffuse Medium
by avoiding such a confusion. The 3.4~$\mu$m profile is therefore
better defined. 

This profile results from identified absorptions of the symmetric and
antisymmetric vibrations of -CH$_2$- (at $\sim$ 3.42 and 3.50~$\mu$m)
and terminal -CH$_3$ (at $\sim$ 3.38 and 3.48~$\mu$m) groups in
aliphatics, and are indicated above optical depth spectra in Fig
\ref{Fig1}. Additional absorption is present around 3.28~$\mu$m, an absorption
generally attributed to aromatic C-H stretching mode.
The resultant optical depth line profiles as seen in NGC~1068,
NGC~7172 and NGC~5506 are similar to the ones observed toward GC IRS7
and GCS3I with the VLT, which implies that the CH$_2$/CH$_3$ aliphatic
component ratio of extragalactic dust is the same in the diffuse
interstellar medium encountered in these objects, i.e about
2$\pm$0.5.
In the case of the Superantennae galaxy (IRAS 19254-7245), the
3.4~$\mu$m is almost saturated ($\tau\sim$0.7) but
\cite{2003ApJ...595L..17R} argue that the profile is similar to our
Galactic center one.

\section{The 3.4~$\mu$m hydrocarbons/10~$\mu$m silicates ratio}

In the analysis of the extragalactic 3.4~$\mu$m, it is common to
calculate the ratio of the 3.4~$\mu$m aliphatic signature of the
diffuse galactic medium to the 10~$\mu$m silicate absorption feature.
This ratio should reflect some differences in the carbon to silicate
abundances at galactic scales. However, for the ratios to be compared
with relevance, one needs an accurate measurement of the scale at
which the extinction in both features are measured
(e.g. \cite{2001ApJ...557..637T}), as well as to take into account the
difference of extension of the infrared background source at 3.4 and
10~$\mu$m. As an example, the comparison of the color
excess derived for NGC7172 by \cite{1997ApJ...477..631V} and the
silicate absorption observed by the IRAS Low Resolution Spectrometer
(\cite{1991MNRAS.248..606R}) are in contradiction when using classical
extinction curves. We used for this particular source a conservative
optical depth for the silicates, derived from the color excess, as the
silicate optical depth of the IRAS LRS spectrum is poorly
constrained. A summary of the 3.4~$\mu$m/9.7~$\mu$m ratios are given in
Table~\ref{Carbone-Silicates}.

Within variations of a factor of two, which could be attributed to
galactic plane and/or inner torus dust inclinations as well as
temperature gradient effects (which for example modify the optical
depth of absorption bands and the scale of the background infrared
emitting source at 3 and 10 $\mu$m), the
C$_{\mbox{Aliphatics}}$/Si$_{\mbox{Silicates}}$ ratio in the DISM, as
probed by the aliphatics and the silicates, is of the same order.

%As pointed out by \cite{2000MNRAS.319..331I},
%__________________________________________________ One column table
   \begin{table}
      \caption[]{Aliphatic hydrocarbons/Silicates ratio}
         \label{Carbone-Silicates}
     $$
         \begin{array}{p{0.15\linewidth}ccc}
            \hline
            \noalign{\smallskip}
{\rm{Source Name}}	&\tau (\mathrm{3.4} \mu \mathrm{m}) 		&\tau (\mathrm{Silicates})    	&\mathrm{\mathrm{3.4} \mu \mathrm{m}/\mathrm{Silicates}} \\
\hline
NGC1068		&0.1\pm0.015		&0.9\pm0.1 ^{\mathrm{a}}    	&0.11\pm0.05 \\
NGC7172		&0.09\pm0.015 		&> 0.3  ^{\mathrm{b}}   	&< 0.35	\\
IRAS19254-7245	&0.7\pm0.1 		&\rm{Saturated}    	&-	\\
NGC5506		&0.04\pm0.01 		&0.9\pm0.2 ^{\mathrm{c}}     	&0.05\pm0.2 	\\
\hline
         \end{array}
     $$
\begin{list}{}{}
\item
%${\mathrm{a}}$) Inclination of the host galaxy, an inclination of 0$^{\mathrm{o}}$ meaning Face on. 
(${\mathrm{a}}$) using the value derived by \cite{2001ApJ...557..637T} in the central 1$''$ of NGC1068. (${\mathrm{b}}$) based on the color excess E(B-V) as explained in \cite{1997ApJ...477..631V}
(${\mathrm{c}}$) Estimated using the Infrared Space Observatory Photometer PHT40 spectrum recorded on 13-Aug-1996. 
\end{list}
   \end{table}

%___________________________________________________

\section{Galactic Center and Extragalactic mid infrared spectra}

\begin{figure*}
   \centering
\begin{minipage}{8.5cm}
    \includegraphics[width=9.cm]{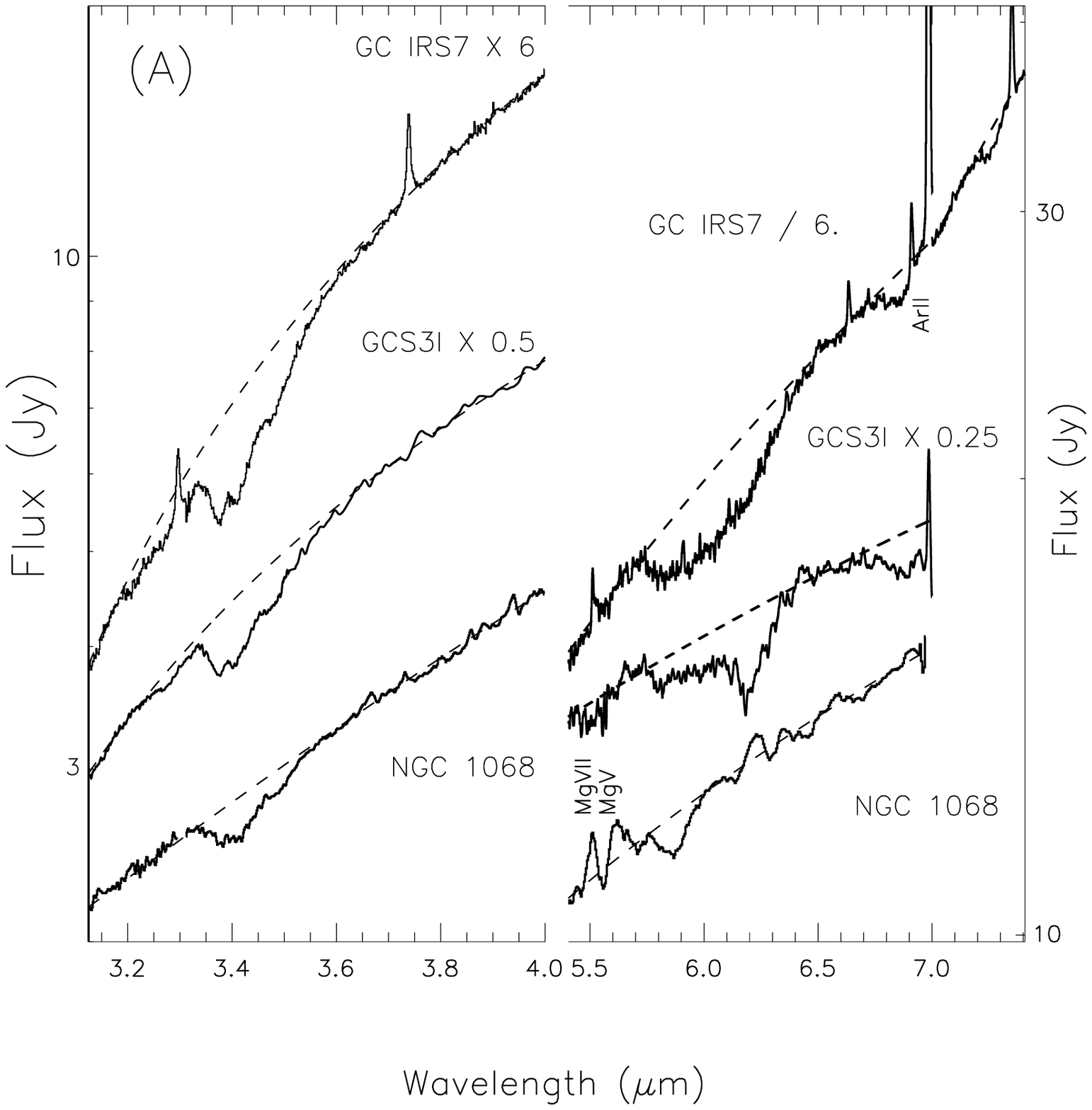}
\end{minipage}
\begin{minipage}{8.5cm}
    \includegraphics[width=9.cm]{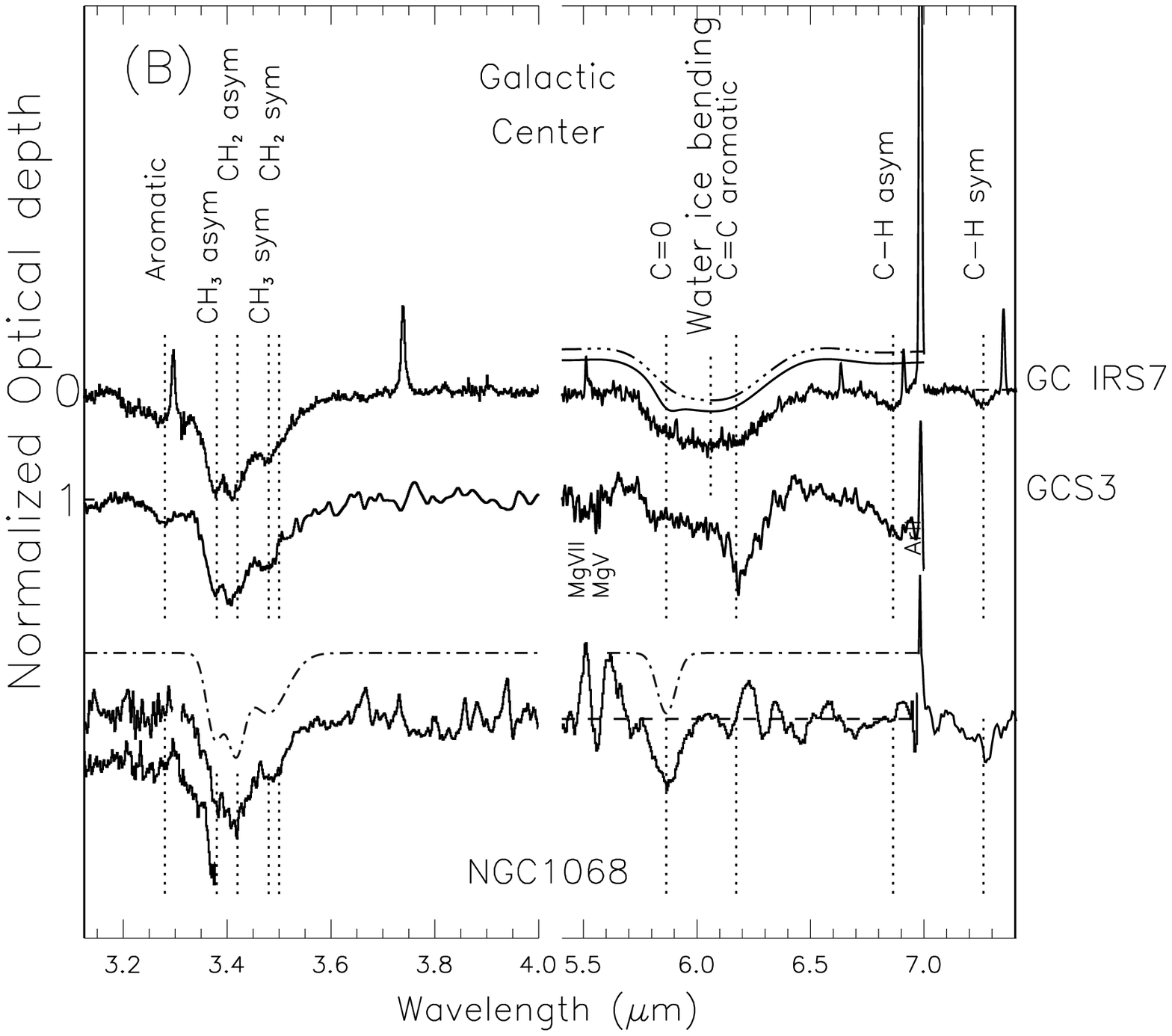}
\end{minipage}
%%%
   \caption{
(A) : Near and mid infrared spectra of two galactic center lines of
sights sampling the diffuse interstellar medium, and one Seyfert~2
inner region. All spectra are scaled as described, for clarity. The
adopted local continua used to establish the optical depth in the
absorption features are shown as dashed lines. (B) : Resultant optical
depth spectra comparing Galactic center dust absorptions and the one
seen along the Seyfert~2 nucleus line of sight of NGC1068. The NIR
(3-4~$\mu$m) spectra are ISO-SWS or VLT-ISAAC spectra, whereas all MIR
(5.5-7.4~$\mu$m) spectra are ISO-SWS ones. The expected position of
caracteristic infrared groups are indicated above. The carbonyl
contribution to the IRS7 spectrum which would remain undetected is
shown above. It is decomposed in the pure water ice OH bending
absorption (dot-dashed) and the maximum undetected carbonyl + OH (full
line). See text for further details.
              \label{Fig2}
		}
%%%
    \end{figure*}

In a review presented by Pendleton and Allamandola (2002) are
summarised the spectra of most of the interstellar analog relevant
material measured in the laboratory, apart from kerogens, and
proposed for an interstellar identification of the 3.4~$\mu$m
absorption features. From this set of data, which includes the long
wavelength region of these materials, it appears clearly that the
simultaneous comparison of the 3-4~$\mu$m range C-H stretching modes
absorption with the 5-8~$\mu$m bending and deformation modes,
``fingerprints'' of the same molecules, now acessible via satellite
observations, will put the most severe constraint on the nature of the
carbonaceous component of the DISM refractive dust.
In order to make such a comparison, and to shed light on the line of
sight confusion/contamination, in Fig \ref{Fig2} we show the
near and mid-infrared spectra of SgR~A$*$/GC IRS$8$, GCS$3$I and
NGC~1068, after subtraction of a local continuum due to
the background infrared emitting region. The overall
optical depth has been normalised to the 3.4~$\mu$m aliphatic region
of the spectrum, in order to directly compare the various absorptions
arising in the whole presented spectra.\\

The mid infrared region of the galactic center individual sources are
quite complicated, due mostly to circumstellar and dense molecular
cloud materials along the lines of sight that do not participate in
the diffuse interstellar medium component.
SgR~A$*$ is dominated in that region by a quite strong water ice
bending mode arising in dense molecular cloud environments. This band
smears out all the very faint relatively large solid state absorptions
that could appear on the continuum of the infrared source. The
decomposition of the spectrum with several absorbing components has
been reported by \cite{2000ApJ...537..749C} but clearly the result is
dependent on the water ice structure and the number of components
used, which are not constrained at all by such a ``Gaussian-like''
absorption band, making the identification very difficult. The fainter
CH$_2$ and CH$_3$ deformation modes are seen at around 6.85~$\mu$m and
7.25~$\mu$m and correspond to what is expected for typical
aliphatic hydrocarbons.
The case of GCS3, presented below and located at only a few arcminutes
from SgR~A$*$, is a good example of the problems encountered in
the identification of mid infrared spectra. The strong 6.2~$\mu$m
absorption feature observed toward this line of sight is in fact of
circumstellar nature, as is observed in several Wolf Rayet
envelopes (\cite{2001ApJ...550L.207C}), which do not display
 any strong 3.4~$\mu$m absorption counterpart.\\

In contrast to the spectra discussed above, the spectra of the nuclear
regions of Seyfert2 Galaxies do not pertain to the same regime and
will not therefore suffer from local circumstellar
contamination. The infrared continuum is thought to be emitted
by a central dust condensation, of a few parsecs of extension
(\cite{2001ApJ...557..637T}, \cite{2000A&A...353..465M}). This offers
us the opportunity to probe the matter with a large infrared beam,
probing essentially the diffuse matter, and lowering the importance of
individual specific objects as they represent a small volume filling
factor along the line of sight.
The spectrum of the NGC1068 nucleus, recorded in the mid infrared by the
ISO short wavelength spectrometer, and displayed in Fig \ref{Fig2} ,
does not show either the 6~$\mu$m water ice bending absorption,
or the circumstellar contribution at 6.2 $\mu$m observed for
GCS3. The spectrum of the nucleus does present a large absorption around
5.87~$\mu$m with an optical depth of about half the 3.4~$\mu$m
one. This absorption, intrinsically very strong in the infrared, is
typical of the C=O carbonyl group and is discussed in relation to
the other signatures in the next section.
%__________________________________________________ One column table
   \begin{table}
      \caption[]{List of detected and/or expected infrared absorption features}
         \label{listeraies}
     $$
         \begin{array}{clc}
            \hline
            \noalign{\smallskip}
\rm{Wavelength}\; (\mu\rm{m})\;\;\;		&\rm{Mode}		&\rm{Note}\\
%\; (\mu\rm{m})		&			&\\
 %$\;$ ($\mu$\rm{m})
\hline
3.28	&\rm{C}$-$\rm{H}\; \rm{aromatic \; stretch.}			&^{\mathrm{a}}\\
3.38	&\rm{CH}_3\; \rm{aliphatic \; asymmetric \; stretch.}			&\\
3.42	&\rm{CH}_2\; \rm{aliphatic \; asymmetric \; stretch.}			&\\
3.48	&\rm{CH}_3\; \rm{aliphatic \; symmetric \; stretch.}			&\\
3.5	&\rm{CH}_2\; \rm{aliphatic \; symmetric \; stretch.}			&\\
5.87	&\rm{C}$=$\rm{O}\; \rm{ketone \; stretch.}				&\\
6.2	&\rm{C}$=$\rm{C}\; \rm{aromatic \; stretch.}			&\\
6.85	&\rm{CH}_2\; \rm{aliphatic \; chain \; like \; deformation \; vibration}	&^{\mathrm{b}}\\
7.25	&\rm{CH}_3 \; \rm{aliphatic \; symmetric \; bending \; deformation}	&\\
9.7	&\rm{SiO} \; \rm{silicates \; stretch.}				&^{\mathrm{c}}\\
\hline
         \end{array}
     $$
\begin{list}{}{}
\item (${\mathrm{a}}$) Possible contribution from =CH stretch. mode. (${\mathrm{b}}$) Principal more intense component, as less intense CH$_3$ asymmetric deformations occur in this range. (${\mathrm{c}}$) Exact position depend on the silicate type.
\end{list}
   \end{table}

%___________________________________________________

 % Two column figure (place early!)
%______________________________________________ 
\section{Constraints on the nature of the hydrocarbons}
%=======
\subsection*{The 3.4~$\mu$m profile}

Given the CH$_2$/CH$_3$ integrated absorption ratio of about two
deduced from the previous analysis, the mean length of the carbon
aliphatic chains (sub-chains if branched aliphatics) encountered both
in the Galactic and Seyfert2 galactic nuclei is therefore of six
carbon atoms if purely aliphatic, and fewer if attached as individual
aliphatic groups to an aromatic structure.

The 3.4~$\mu$m absorption profile is to a first order compatible with
many materials (see for example the Table 3 of Pendleton \&
Allamandola, 2002). The only firm constraints obtained from these
absorption bands are : (i) an estimate of the CH$_3$ to CH$_2$ ratio
in the aliphatic component of the ISM dust can be made, (ii) aliphatic
chains are interconnected or branched, leading to broad vibrational
profiles, as simple aliphatics such as the ones presented in the
appendix would display much sharper transitions, and (iii) the
aromatic contribution, through its C-H stretching component around
3.28~$\mu$m is at most as abundant as the aliphatic one, except if
these aromatics are very large or dehydrogenated.\\ In addition to the
last statement, the 6.2~$\mu$m aromatic C=C vibration is absent or
weak from the Galactic center spectra, whereas this mode is much
stronger than the aromatic CH stretch, which implies there is at most
a small contribution of aromatics to this DISM material.
%=======
%%%%
\subsection*{Comparison with the carbonaceous phases in meteorites}
The spectra of the insoluble phase of carbonaceous chondrites, such as
Orgueil and Murchison, are often considered as excellent candidates
analogs for the ISM carbonaceous component, based on a good fit of the
3.4~$\mu$m of its aliphatic phase (\cite{1991A&A...252..712E},
\cite{1995P&SS...43.1359P}). However, the Orgueil and Murchison insoluble
material possess a C/H ratio of about 17 (\cite{2000E&PSL.184....9G}),
which shows that this extracted material is not dominated by aliphatic
chains, but rather by aromatic material, as already observed
(\cite{2000E&PSL.184....9G}, \cite{1993GeCoA..57..933D}). If aliphatic
chains were dominant, the CH$_2$/CH$_3$ network would imply a C/H
ratio much lower, closer to 0.5. The long wavelength absorptions in
the infrared spectra of these meteorites extracts are a mixture of the
aliphatics deformation modes and of the bulk of the aromatics
present. When the sample is heated above about 600-700K under vacuum,
the absorptions seen at 3.4~$\mu$m disappear whereas most of the
strongest absorptions in the 5 to 10~$\mu$m range remains at higher
temperatures (\cite{1988ApJ...328L..75W}), demonstrating that they are due
to at least two different phases, the less volatile being attributed
to large polyaromatics. This explains why parts of the Orgeuil and
Murchison meteorites are able to reproduce the interstellar 3.4~$\mu$m
profile but fail to account for the now observed longer wavelength
part of the spectrum.
%======
\subsection*{Oxygen content of the DISM aliphatics}
As discussed previously, the NGC1068 mid infrared spectrum display a
carbonyl absorption at 5.87~$\mu$m. The absorption peak position
observed is caracteristic of ketones (R-(C=O)-R') or carboxylic acids
(R-(C=O)-O-H). The absence of a strong carboxylic (O-H) feature in the
Seyfert~2 spectra, is even more specific to the ketones.
Using a decomposition with Gaussians as shown in Fig \ref{Fig2} of
both the 3.4 and 5.87~$\mu$m integrated absorbance, we can estimate
the C/O ratio needed to account for this new absorption band. Using
$A$(CH$_3$)$=$1.25$\times$10$^{-17}$ cm.group$^{-1}$ and
$A$(C=0)$=$2.2$\times$10$^{-17}$ cm.group$^{-1}$ (see Appendix where
we determine experimentally the integrated absorbances of ketone
carbonyl and CH stretches; see also \cite{1986A&AS...64..453D} )
together with the ratio of integrated absorbance deduced from the fit,
we obtain a CH$_3$/CO ratio of about 3. It leads to a C/O of
at least 9 taking into account a CH$_2$/CH$_3$ ratio of about
2.

The fact we do not see large amounts of oxygen inserted in the
aliphatic chains is not so surprising given the oxygen budget in the
solid phase of the diffuse medium, where a large part is already
locked in the silicates. Indeed, if we use the integrated absorbance
values of $A$(Si-O)$\approx$1.6$\times$10$^{-16}$ cm.molecule$^{-1}$
(\cite{1998dartois}) for the stretching mode of pyroxenes, $\Delta
\bar{\nu} \approx 300$cm$^{-1}$ for the FWHM of the band,
A$_{\rm{V}}$/$\tau_{\rm{(Silicates@9.7\mu\rm{m})}}$$\sim$16.6 or 18.5
(\cite{1985ApJ...288..618R}; \cite{1990ARA&A..28...37M}),
N$_{\rm{H}}$/A$_{\rm{V}}$$\approx$1.9$\times$10$^{21}$cm$^{-2}$.mag$^{-1}$
(\cite{1978ApJ...224..132B}), and given that there are 3 oxygen atom
per Si in a pyroxene (the most common pyroxene form is
(Ca,Mg,Fe)$_2$Si$_2$O$_6$), we can
deduce~O/H$\approx$1.7$\times$10$^{-4}$. Using a value of the oxygen
atomic abundance of 450 to 550 ppm (\cite{2001ApJ...554L.221S}), about
30-40\% of the oxygen is locked in the silicates. A large part of
the oxygen seems also confined to the gas phase, in the atomic phase
even at moderate to high extinctions (\cite{2003ApJ...591.1000A},
\cite{2003AAS...203.7003B}).\\

In infrared spectroscopy, the absence of a line in a spectrum for an
infrared active group with a high integrated absorption coefficient
(such as the carbonyl) is often a very strong constraint to define the
nature of the material, in relation to the detected bands. The
appearance of a carbonyl absorption band in the Galactic center lines
of sight should not be affected by the water ice observed one, as the
former lies in the foreground. This means we can put a useful upper
limit on the oxygen content. Based on the mid infrared spectrum of
SgR~A$*$ and using the same carbonyl profile as the one used for
NGC1068, we estimate the maximum O/C that would escape detection,
as shown in Fig. \ref{Fig2}. The upper limit derived in this way
translates into an C(aliphatic)/O(carbonyl) $\leq$40.\\

Absence of the ketone carbonyl toward galactic center sources, besides
the strength of this infrared absorption, suggests that the synthesis
of the carbonaceous ISM is formed in a very reducing environment such
as the one shown by \cite{2002AdSpR..30.1451M}. This is a strong
constraint as, for example, almost all the laboratory spectra presented by
\cite{2002ApJS..138...75P} present a carbonyl absorption, even the
ones in which no oxygen containing molecules were introduced in the
experiment, but rather resulted from pollution, showing the ability of
oxygen to insert in the aliphatic network and the ease of measuring
this particular carbonyl mode.

In summary, observations shown in this paper tend to show that the
aliphatic component of Galactic dust and the one encountered toward
the nucleus of NGC~1068 are globally the same, except for the possible
presence of a small amount of ketones. This last finding should of course
be confirmed by further observations at mid infrared wavelength of the
Seyfert~2 central dust condensation.

%======

\subsection*{Cycling, timescales and column densities constraints}

%The aliphatics responsible for the 3.4~$\mu$m absorption band resist
%quite efficiently to the DISM UV field. This is an indication that
%this material is not made of simple hydrocarbons with a CH$_2$/CH$_3$
%ratio of 2, such as hexane, otherwise it would be destroyed in a quite
%short timescale, as discussed by \cite{2001A&A...367..347M}.

Until now, the observations toward the dense medium do not show the
presence of any dense cloud 3.4~$\mu$m absorption, whereas the column densities
are large enough to detect it at levels comparable or better than the
diffuse medium. Laboratory ice analogues are able to produce
refractory materials, stable after the evaporation of the ices, with
an efficiency of at most 10$^{-2}$ of the initial total ice mass
content (e.g. \cite{2003A&A...412..121M} and references therein).
Considering an optimistic maximum total ice abundance ratio of a few
10$^{-4}$, at most 10$^{-6}$ of the total material, with respect to
the hydrogen density, is processed in this way in one cycle of a dense
cloud. The observed diffuse medium
C$_{\mbox{Aliphatics}}$/C$_{\mbox{Total}}$ abundance is around 5 to
10\% (\cite{1994ApJ...437..683P}, \cite{1991ApJ...371..607S}). One
such cycle can therefore produce at most one tenth of the observed DISM
aliphatics. 

A grain cycling from diffuse to dense medium of at least ten times is
required to explain the observations, if we suppose the produced
residues are not destroyed and remain in the subsequent cycling
sequence. This finding is in contradiction to the fact we do not see
the 3.4~$\mu$m C-H absorptions in at least some of the observed dense
clouds.

An alternative interesting scenario would be that in the diffuse
medium, due to the high atomic hydrogen content in the gas phase, there exists
a mechanism that can re-hydrogenate the aliphatic carbon backbone of
the otherwise hydrogen UV abstracted material
(\cite{2002AdSpR..30.1451M}, \cite{2001A&A...367..347M},
\cite{2001A&A...367..355M}). When entering the dense clouds, the
atomic hydrogenation is stopped, due to the passage of a reducing
environment dominated by atomic hydrogen to a rather unreactive
molecular hydrogen one, and giving rise to a chemistry leading to the
formation of the ice mantles. The CH stretching modes would then
disappear. To validate this scenario, one must therefore search for
the mid infrared signature, if strong enough, of the much more stable
carbon backbone still present, in deeply embedded objects.
\subsection*{Galactic versus extragalactic 3.4~$\mu$m}
In our Galaxy, the 3.4 $\mu$m absorption feature is sensitive to the
environment, as it is not observed in dark clouds. Seyfert~2 and
Galactic center lines of sights should probe different mediums, but
the solid absorption features in the type 2 Active Galactic Nuclei
lines of sight and our Galactic diffuse medium share common properties
as underlined by the similarities of their carbonaceous infrared
absorption bands.

AGNs possess a powerful central source (Black hole + accretion disk)
that produces the ionizing radiation that heats the dust we see. The
temperature of the dust is shown to decrease very fast with the
distance to the central source, to reach 100-200K at a few hundred
parsecs. At the location where the bulk of the absorption takes place,
the temperature might not differ too much from the one encountered in
the diffuse medium. The material seen via these features must be very
UV-resistant and in the solid state as it must stand the strong UV
field of the Diffuse Medium. Therefore, only if the temperature of
these carbonaceous grains could reach above about 300K would we see
these lines in emission. The case of the 3.3~$\mu$m, the so called PAH
feature, is different as the excitation mechanism is molecular and
transient after absorption of single UV photons by relatively small
molecular systems.

An important clue toward the understanding of the 3.4~$\mu$m feature
composition lies in the dust cycling timescales of the inner dust
torus of Seyfert~2 versus our Galactic diffuse medium. They should be
quite different. It seems to favor an interpretation of in-situ
formation of this 3.4~$\mu$m component, without invoking many dust
cycles, or the hydrogenation of an almost pure carbon dominated dust,
as explained above.
%

%======

\section{Conclusions}

   \begin{itemize}

      \item The profiles of the aliphatic 3.4~$\mu$m absorption
      features measured in four Seyfert~2 nuclei lines of sight are
      similar to those in our Galaxy, suggesting that both the length
      of carbon chains and the nature of the CH stretching modes are
      similar in these objects.

      \item The material responsible for our galactic diffuse
      interstellar medium absorption bands is a relatively simple
      network containing aliphatics linked together and/or to aromatic
      structure, and contain little oxygen, as no strong
      bands arising from other strong infrared active groups are seen
      in the Mid-Infrared spectra.

      \item A hint for a carbonyl absorption in the line of sight
      toward the nucleus of the Seyfert~2 galaxy NGC~1068 is
      reported. This absorption should be confirmed by further
      satellite investigations. We recall that use of mid-IR medium
      resolution spectrometers onboard next generation satellites
      is the only way to get physical insight into the nature of solid
      state refractory matter. The almost featureless continuum
      spectra of some Seyfert~2 nuclei provide excellent lines of
      sight to search for the mid-IR signatures of the aliphatics seen
      at 3.4~$\mu$m, an often difficult task for our Galactic center
      specific lines of sights, due to local circumstellar
      contamination. Spectra obtained at resolution lower than about
      1000 in the mid infrared are of little interest with respect to
      the exact spectroscopic identification of solid matter composition.

      \item The numerous extragalactic spectra recorded by the Spitzer
      Space Telescope will however allow us to obtain the spatial
      distribution of the DISM solid aliphatic mid-infrared features,
      and to assess their importance in the cycle of dust at large
      scales.

   \end{itemize}

\begin{acknowledgements}
      Part of this work was funded and performed during a visiting
      scientist program of one month at ESO Chile Headquarters. The
      authors wish to thanks the anonymous referee as well as
      G. Matrajt for fruitful comments that improved the submitted
      paper.
\end{acknowledgements}

\section*{Appendix: Integrated absorbances of ketones and alcanes at 10K}
%______________________________________________ 
   \begin{figure*}
%   \centering
\begin{minipage}{8cm}
    \includegraphics[width=7cm,angle=90]{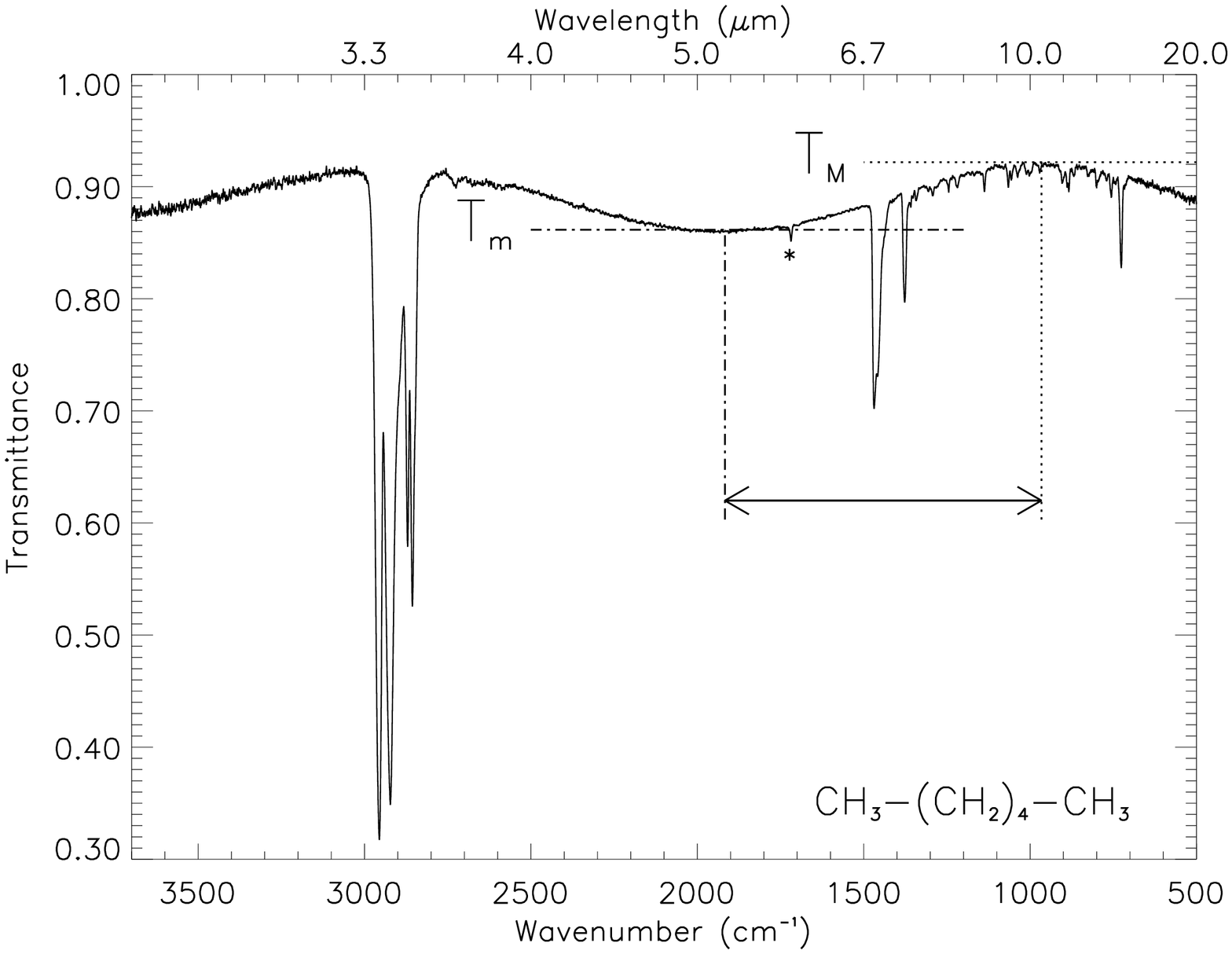}
\end{minipage}
\begin{minipage}{1cm}
\hspace*{1cm}
\end{minipage}
\begin{minipage}{8cm}
    \includegraphics[width=7cm,angle=90]{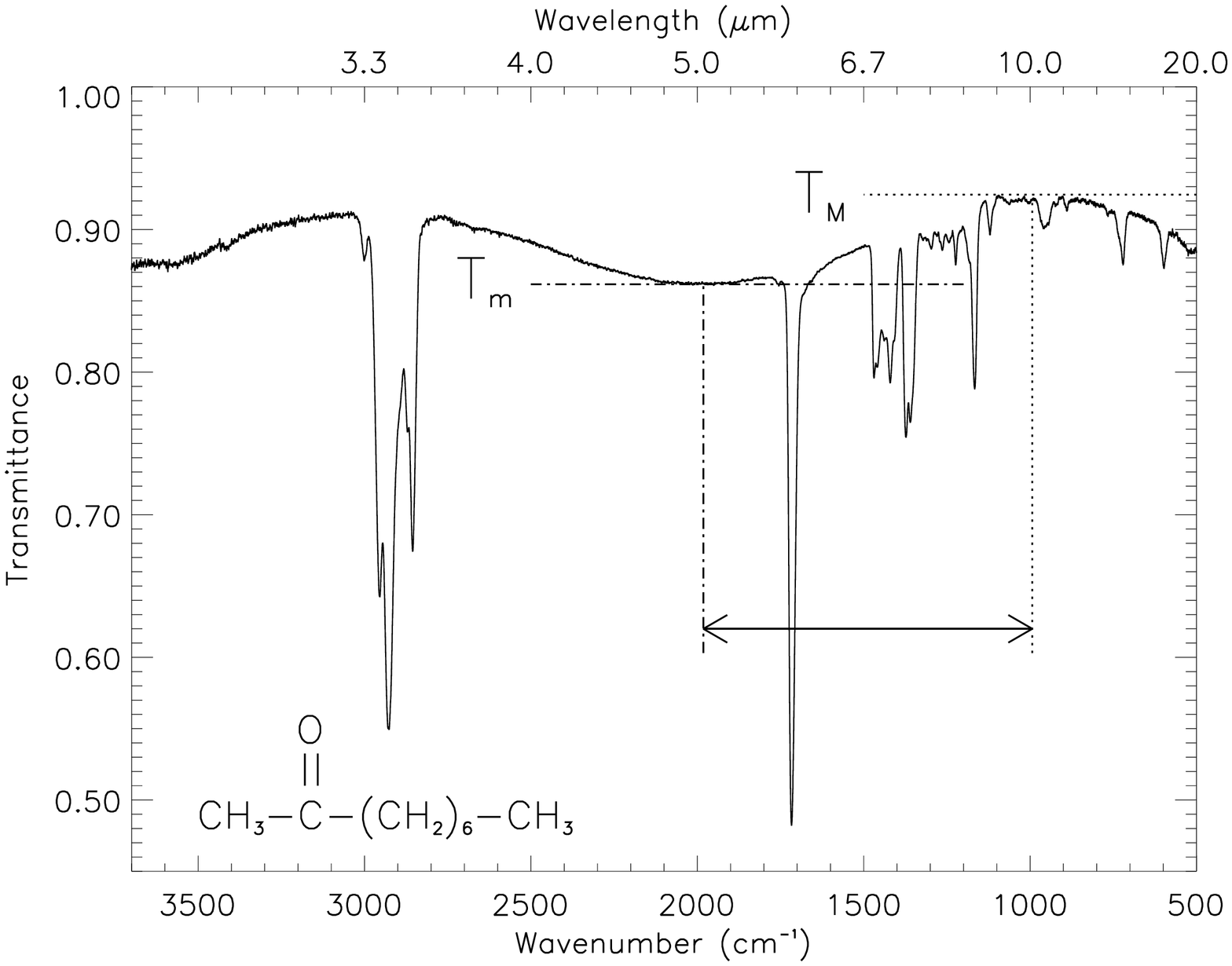}
\end{minipage}
   \caption{Transmittance spectra of a film of hexane ice (left) and
   2-nonanone ice (right) deposited on a thick CsI window, at 10K. Note the
   fringes due to the plane parallel ice film acting as an infrared
   Fabry-Perot. The meaning and use of the local maxima T$_{\rm{M}}$ and
   T$_{\rm{m}}$ as well as their frequency separation indicated on the
   plot are explained in the text.
              \label{franges}%
		}
    \end{figure*}
%______________________________________________ 
The transmittance T of a plane-parallel film deposited on a non-absorbing thick substrate is a complex function of the wavenumber $\bar{\nu}$, the substrate refractive index n$_{\rm{S}}$, the film refractive index n$_{\rm{F}}$, the imaginary part of the film index k$_{\rm{F}}$, and the film thickness d$_{\rm{F}}$ (\cite{Mini}). In the regions of the spectrum where absorption is very weak (k$_{\rm{F}}<<$1), this equation takes the simpler form :
$$
\rm{T}=\frac{\rm{A}}{\rm{B}-\rm{C cos}(\phi)+\rm{D}}
$$
where A$=16$n$_{\rm{F}}^2$n$_{\rm{S}}$, 
B$=($n$_{\rm{F}}+1)^3$ $($n$_{\rm{F}}+$n$_{\rm{S}}^2)$,
C$=2($n$_{\rm{F}}^2-1)$ $($n$_{\rm{F}}^2-$n$_{\rm{S}}^2)$,
D$=($n$_{\rm{F}}-1)^3$ $($n$_{\rm{F}}-$n$_{\rm{S}}^2)$,
$\phi=$4$\pi$n$_{\rm{F}}$$\bar{\nu}$d$_{\rm{F}}$.\\
%x$=$exp$(-$4$\pi$k$_{\rm{F}}$$\bar{\nu}$d$_{\rm{F}}$).
This transmittance shows a fringe pattern whose maximum and minimum are given, in the case of n$_{\rm{F}}$$<$n$_{\rm{S}}$, by:
$$
\rm{T}_{\rm{M}}=\frac{\rm{A}}{\rm{B}+\rm{C}+\rm{D}}
\; \rm{and} \; \rm{T}_{\rm{m}}=\frac{\rm{A}}{\rm{B}-\rm{C}+\rm{D}}
$$ 
Using the measured transmittance difference between two adjacent
maxima, one can therefore obtain the refractive index in a clean
(k$_{\rm{F}}<<$1) region of the spectrum by rearranging the above equations in
the form : 
$$
\rm{n}_{\rm{F}}=\sqrt{\frac{\gamma \pm \sqrt{\gamma^2-4\rm{n}_{\rm{S}}^2}}{2}} \;\; \rm{where} \;\; \gamma=(1+\rm{n}_{\rm{S}}^2+4\rm{n}_{\rm{S}}(1/T_{M}-1/T_{m}))
$$

We apply this property of the transmittance to the infrared spectrum
of hexane deposited at 10K on a CsI substrate, as shown in Fig
\ref{franges}. The T$_{\rm{m}}$ and T$_{\rm{M}}$ are evaluated around
1900 and 1000cm$^{-1}$, respectively.
The CsI substrate refractive index n$_{\rm{S}}\sim$1.75 at cryogenic
temperatures. We therefore deduce that n$_{\rm{F}}\sim$1.41 using the
above equation. 

The thickness can now be evaluated using the wavenumber difference
between maxima, as presented in Fig \ref{franges}, using the classical
interfringe relation $\rm{d}_{\rm{F}}\rm{\;(cm)}=1/[2 \Delta \bar{\nu}\rm{\;(cm}^{-1}\rm{)}
\rm\;{n}_{\rm{F}}$], where $\Delta \bar{\nu}$ is the wavenumber
difference between two adjacent maxima (minima or maxima) of the
fringes.  The film presented in Fig \ref{franges} has therefore a
thickness of 1.86~$\mu$m.

\begin{figure}
%   \begin{minipage}{10cm}
%   \includegraphics[height=10cm,angle=90]{hexane_3_4.eps}
    \includegraphics[height=10cm,angle=90]{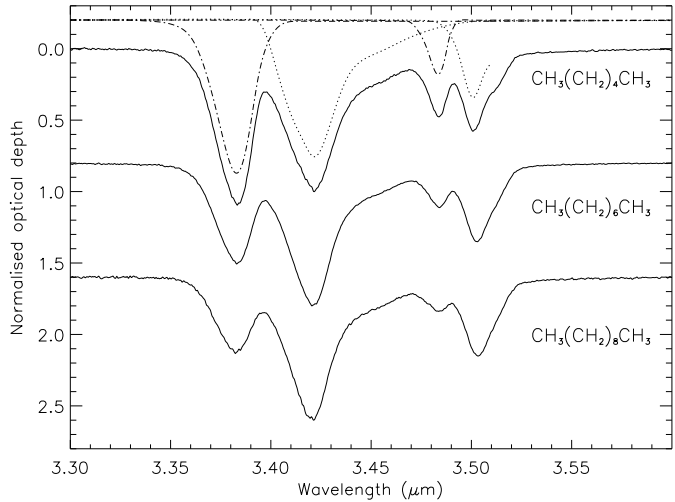}
%   \end{minipage}
%   \begin{minipage}{7cm}
   \caption{
Optical depth spectra of hexane, octane and decane at 10K in the CH
stretch infrared region of the spectrum, after fringes local baseline
correction. All spectra have been normalized to the antisymmetrical
CH$_2$ stretching mode around 3.42~$\mu$m for direct comparison of the
profiles, and shifted for clarity. The hexane individual stretching
modes profiles are displayed in the upper part of the figure. The
CH$_3$ and CH$_2$ stretching modes relative ratios are directly
proportionnal to the relative contents of the alkanes shown here.
              \label{alcanes_3_4}%
		}
%   \end{minipage}
    \end{figure}

Assuming that n$_{\rm{F}}$ does not vary too much in an alkane
absorption band, one can now evaluate the integrated absorbance of
the infrared active transitions. We subtract a local baseline by
modelling the fringes pattern around an absorption and extract the
optical depth $\tau$ of the band. The integrated absorbance of a line 
is now related to this $\tau$ by :
$$
\rm{{\it A} (cm/molecule)} = \int_{line} \frac{\tau_{\bar{\nu}} \rm{M}} {\rm{d}_{\rm{F}} \; \rm{\it N}_A \; \rho}\;d\bar{\nu}
$$
$$
\;\; \rm{and} \;\; 
\rm{{\it A} (cm/group)} = \frac{\rm{{\it A} (cm/molecule)}}{\rm{N}_g}
$$
where M is the molar mass, d$_{\rm{F}}$ the film thickness, {\it
N}$_A$ the Avogadro number, $\rho$ the density of the ice and N$_g$
the number of equivalent vibrational groups of the molecule implied in the transition.
Using the spectra recorded at our institute at 10K of hexane, octane
and decane, and after comparison of the baseline subtracted spectra,
one can deduce the individual profiles for the various CH$_2$ and
CH$_3$ stretching modes, as shown in Fig \ref{alcanes_3_4}, where
they are normalized to the CH$_2$ antisymmetrical stretch around 2920
cm$^{-1}$. Using a density of 0.68~g.cm$^{-1}$ for the hexane ice, we
estimate {\it A}(d-CH$_3$)=1.25$\pm$0.1$\times$10$^{-17}$ cm.group$^{-1}$ at 3.38~$\mu$m,
{\it A}(a-CH$_2$)=8.4$\pm$0.1$\times$10$^{-18}$ cm.group$^{-1}$ at 3.42~$\mu$m,
{\it A}(s-CH$_3$)=2.0$\pm$0.1$\times$10$^{-18}$ cm.group$^{-1}$ at 3.47~$\mu$m,
{\it A}(s-CH$_2$)=2.4$\pm$0.1$\times$10$^{-18}$ cm.group$^{-1}$ at 3.50~$\mu$m.

%

%______________________________________________ 
%______________________________________________ 

%______________________________________________ 
   \begin{figure}
%   \begin{minipage}{10cm}
%   \includegraphics[height=10cm,angle=90]{heptanone_3_4.eps}
    \includegraphics[height=10cm,angle=90]{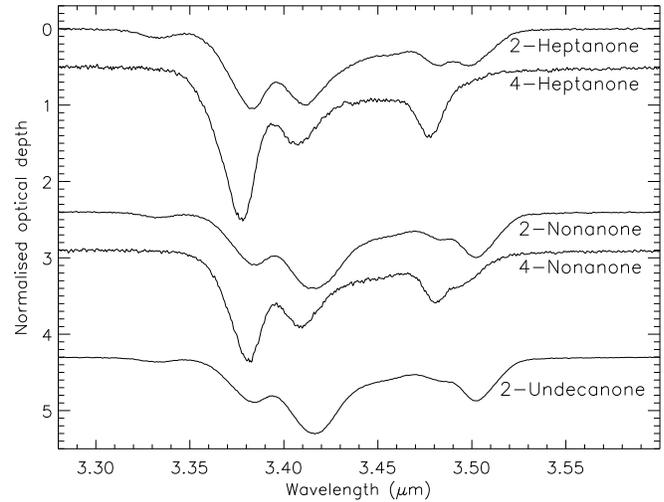}
%   \end{minipage}
%   \begin{minipage}{7cm}
   \caption{
Optical depth spectra of heptanone, nonanone and undecanone where the
carbonyl is either positionned at the end or the middle of the carbon
skeleton, after fringes local baseline correction, recorded at 10K in
the CH stretch infrared region of the spectrum. All spectra have been
normalized to the antisymmetrical CH$_2$ stretching mode around
3.42~$\mu$m for direct comparison of the profiles, and shifted for
clarity.
              \label{cetones_3_4}%
		}
%   \end{minipage}
    \end{figure}
%______________________________________________ 

%______________________________________________ 
   \begin{figure}
%   \begin{minipage}{10cm}
%   \includegraphics[height=10cm,angle=90]{heptanone_6_5.eps}
   \includegraphics[height=10cm,angle=90]{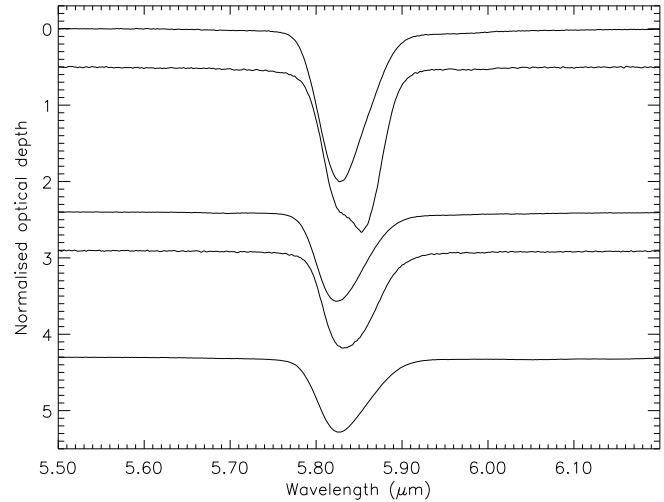}
%   \end{minipage}
%   \begin{minipage}{7cm}
   \caption{
Optical depth spectra of heptanone, nonanone and undecanone, where the
carbonyl is either positionned at the end or the middle of the carbon
skeleton, recorded at 10K in the C=O stretch infrared region of the
spectrum. All spectra have been normalized to the antisymmetrical
CH$_2$ stretching mode around 3.42~$\mu$m for direct comparison of the
profiles, and shifted for clarity.
              \label{cetones_6_5}%
		}
%   \end{minipage}
    \end{figure}
%______________________________________________ 


\begin{thebibliography}{}

\bibitem[Allen \& Wickramasinghe(1981)]{1981Natur.294..239A} Allen, 
D.~A.~\& Wickramasinghe, D.~T.\ 1981, \nat, 294, 239 

\bibitem[Andr{\' e} et al.(2003)]{2003ApJ...591.1000A} Andr{\' e}, M.~K.~et 
al.\ 2003, \apj, 591, 1000 

\bibitem[Baumgartner(2003)]{2003AAS...203.7003B} Baumgartner, W.~H.\ 2003, 
American Astronomical Society Meeting, 203.

\bibitem[Bohlin, Savage \& Drake(1978)]{1978ApJ...224..132B} Bohlin, 
R.~C., Savage, B.~D., \& Drake, J.~F.\ 1978, \apj, 224, 132 

\bibitem[Bonatto \& Pastoriza(1997)]{1997ApJ...486..132B} 
Bonatto, C.~J.~\& Pastoriza, M.~G.\ 1997, \apj, 486, 132 

\bibitem[Chatzichristou(2000)]{2000ApJS..131...71C} Chatzichristou, E.~T.\ 
2000, \apjs, 131, 71

\bibitem[Chiar et al.(2002)]{2002ApJ...570..198C} Chiar, J.~E., Adamson,
A.~J., Pendleton, Y.~J., Whittet, D.~C.~B., Caldwell, D.~A., \& Gibb,
E.~L.\ 2002, \apj, 570, 198

\bibitem[Chiar \& Tielens(2001)]{2001ApJ...550L.207C} Chiar, J.~E.~\&
Tielens, A.~G.~G.~M.\ 2001, \apjl, 550, L207

\bibitem[Chiar et al.(2000)]{2000ApJ...537..749C} Chiar, J.~E., Tielens,
A.~G.~G.~M., Whittet, D.~C.~B., Schutte, W.~A., Boogert, A.~C.~A., Lutz,
D., van Dishoeck, E.~F., \& Bernstein, M.~P.\ 2000, \apj, 537, 749

\bibitem[Dartois et al.(2002)]{2002A&A...394.1057D} Dartois, E.,
d'Hendecourt, L., Thi, W., Pontoppidan, K.~M., \& van Dishoeck, E.~F.\
2002, \aap, 394, 1057

\bibitem[Dartois 1998]{1998dartois} Dartois, E.,
Thesis, Universit\'e Paris VI, 1998.

\bibitem[Ehrenfreund et al. (1991)]{1991A&A...252..712E} Ehrenfreund, P., Robert, F., 
D'Hendencourt, L., \& Behar, F.\ 1991, \aap, 252, 712 

\bibitem[Eisenhauer et al.(2003)]{2003Msngr.113...17E} Eisenhauer, F.~et 
al.\ 2003, The Messenger, 113, 17 

\bibitem[Gardinier et al.(2000)]{2000E&PSL.184....9G} Gardinier, A., 
Derenne, S., Robert, F., Behar, F., Largeau, C., \& Maquet, J.\ 2000, Earth 
and Planetary Science Letters, 184, 9 

\bibitem[d'Hendecourt \& Allamandola(1986)]{1986A&AS...64..453D}
Dhendecourt, L.~B.~\& Allamandola, L.~J.\ 1986, \aaps, 64, 453

\bibitem[Hickson (1994)]{1994hickson}
Hickson, P., Atlas of compact groups of galaxies, 1994, Gordon and
Breach Science Publishers S.A..

\bibitem[Imanishi(2002)]{2002ApJ...569...44I} Imanishi, M.\ 2002, \apj,
569, 44

\bibitem[Imanishi \& Dudley(2000)]{2000ApJ...545..701I} Imanishi, M.~\&
Dudley, C.~C.\ 2000, \apj, 545, 701

\bibitem[Imanishi(2000)]{2000MNRAS.319..331I} Imanishi, M.\ 2000, \mnras,
319, 331

\bibitem[Ishii, Nagata, Chrysostomou \& Hough(2002)]{2002AJ....124.2790I}
Ishii, M., Nagata, T., Chrysostomou, A., \& Hough, J.~H.\ 2002, \aj, 124,
2790

\bibitem[Jarrett et al.(2003)]{2003AJ....125..525J} Jarrett, T.~H., 
Chester, T., Cutri, R., Schneider, S.~E., \& Huchra, J.~P.\ 2003, \aj, 125, 
525

\bibitem[Laureijs et al.(2000)]{2000A&A...359..900L} Laureijs, R.~J.~et
al.\ 2000, \aap, 359, 900

\bibitem[Leger et al. (1989)]{1989irsa.rept..189L} Leger, A., Dhendencourt, L.~B., 
Verstraete, L., \& Ehrenfreund, P.\ 1989, Infrared Spectroscopy in 
Astronomy, 189 

\bibitem[Leger \& Puget(1984)]{1984A&A...137L...5L} Leger, A.~\& Puget, 
J.~L.\ 1984, \aap, 137, L5

\bibitem[Malkan, Gorjian \& Tam(1998)]{1998ApJS..117...25M} Malkan, M.~A., 
Gorjian, V., \& Tam, R.\ 1998, \apjs, 117, 25 

\bibitem[Marco \& Alloin(2000)]{2000A&A...353..465M} Marco, O.~\& Alloin,
D.\ 2000, \aap, 353, 465

\bibitem[Mathis(1990)]{1990ARA&A..28...37M} Mathis, J.~S.\ 1990, \araa, 28, 
37

\bibitem[Mennella et al. (2002)]{2002AdSpR..30.1451M} Mennella, V., Brucato, J.~R., 
Colangeli, L., \& Palumbo, P.\ 2002, Advances in Space Research, 30, 1451 

\bibitem[Mennella et al.(2001)]{2001A&A...367..355M} Mennella, V., Mu{\~ 
n}oz Caro, G.~M., Ruiterkamp, R., Schutte, W.~A., Greenberg, J.~M., 
Brucato, J.~R., \& Colangeli, L.\ 2001, \aap, 367, 355 

\bibitem[Mini (1982)]{Mini} Mini, A.\ 1982, Thesis, Grenoble.

\bibitem[Mu{\~ n}oz Caro \& Schutte(2003)]{2003A&A...412..121M} Mu{\~ n}oz 
Caro, G.~M.~\& Schutte, W.~A.\ 2003, \aap, 412, 121 

\bibitem[Mu{\~ n}oz Caro et al.(2001)]{2001A&A...367..347M} Mu{\~ n}oz 
Caro, G.~M., Ruiterkamp, R., Schutte, W.~A., Greenberg, J.~M., \& Mennella, 
V.\ 2001, \aap, 367, 347 

\bibitem[Pendleton \& Allamandola(2002)]{2002ApJS..138...75P} Pendleton,
Y.~J.~\& Allamandola, L.~J.\ 2002, \apjs, 138, 75

\bibitem[Pendleton(1995)]{1995P&SS...43.1359P} Pendleton, Y.~J.\ 1995, 
\planss, 43, 1359 

\bibitem[Pendleton et al.(1994)]{1994ApJ...437..683P} Pendleton, Y.~J.,
Sandford, S.~A., Allamandola, L.~J., Tielens, A.~G.~G.~M., \& Sellgren, K.\
1994, \apj, 437, 683

\bibitem[Rieke \& Lebofsky(1985)]{1985ApJ...288..618R} Rieke, G.~H.~\&
Lebofsky, M.~J.\ 1985, \apj, 288, 618

\bibitem[Rigopoulou et al.(1999)]{1999AJ....118.2625R} Rigopoulou, D., 
Spoon, H.~W.~W., Genzel, R., Lutz, D., Moorwood, A.~F.~M., \& Tran, Q.~D.\ 
1999, \aj, 118, 2625 

\bibitem[Risaliti et al.(2003)]{2003ApJ...595L..17R} Risaliti, G.~et al.\ 
2003, \apjl, 595, L17 

\bibitem[Roche et al. (1991)]{1991MNRAS.248..606R} Roche, 
P.~F., Aitken, D.~K., Smith, C.~H., \& Ward, M.~J.\ 1991, \mnras, 248, 606 

\bibitem[Sandford et al.(1991)]{1991ApJ...371..607S} Sandford, S.~A.,
Allamandola, L.~J., Tielens, A.~G.~G.~M., Sellgren, K., Tapia, M., \&
Pendleton, Y.\ 1991, \apj, 371, 607

\bibitem[Sofia \& Meyer(2001)]{2001ApJ...554L.221S} Sofia, U.~J.~\& Meyer, 
D.~M.\ 2001, \apjl, 554, L221 

\bibitem[Spoon et al.(2002)]{2002A&A...385.1022S} Spoon, H.~W.~W., Keane,
J.~V., Tielens, A.~G.~G.~M., Lutz, D., Moorwood, A.~F.~M., \& Laurent, O.\
2002, \aap, 385, 1022

\bibitem[Tomono et al. (2001)]{2001ApJ...557..637T}
Tomono, D., Doi, Y., Usuda, T., \& Nishimura, T.\ 2001, \apj, 557, 637

\bibitem[de Vries et al.(1993)]{1993GeCoA..57..933D} de Vries, M.~S., 
Reihs, K., Wendt, H.~R., Golden, W.~G., Hunziker, H.~E., Fleming, R., 
Peterson, E., \& Chang, S.\ 1993, \gca, 57, 933 

\bibitem[Veilleux, Goodrich \& Hill(1997)]{1997ApJ...477..631V} Veilleux, 
S., Goodrich, R.~W., \& Hill, G.~J.\ 1997, \apj, 477, 631 

\bibitem[Wdowiak, Flickinger \& Cronin(1988)]{1988ApJ...328L..75W} 
Wdowiak, T.~J., Flickinger, G.~C., \& Cronin, J.~R.\ 1988, \apjl, 328, L75 

\end{thebibliography}
\end{document}